\documentclass[aps,prl,groupedaddress,twocolumn]{revtex4}
\usepackage{graphicx}

\begin{document}

\title{Transverse excitations and zigzag transition in quasi-1D hard-disk system
}

\author{A. Huerta$^1$\footnote[1]{Sabbatical leave at Instituto de F\'isica, Universidad Nacional Aut\'onoma de M\'exico (UNAM), 
		Apartado Postal 20-364, 01000 M\'exico, Distrito Federal, M\'exico}, T. Bryk$^{2,3}$,  A. Trokhymchuk$^{2}$}
\affiliation{
$^1$ Facultad de F\'isica, Universidad Veracruzana, Circuito Gonz\'alo Aguirre Beltr\'an s/n Zona Universitaria, Xalapa, Veracruz, C.P. 91000, M\'exico\\
$^2$Institute for Condensed Matter Physics of the
National Academy of Sciences of Ukraine,
1 Svientsitskii Street, UA-79011 Lviv, Ukraine\\
$^3$ Institute of Applied Mathematics and Fundamental Sciences, Lviv Polytechnic National University, UA-79013 Lviv, Ukraine
}

\date{\today}

\begin{abstract}
 Molecular dynamics computer simulations of collective excitations in a system of hard disks confined to a narrow channel of the specific width, that  resembles 2D triangular lattice 
 at disk close packing, are performed. 
 We found that transverse excitations, which for hard-disk system are absent in the limit of 1D and are of acoustic nature  in the limit of 2D, in the case of q1D hard-disk system emerge 
 in the form of transverse optical excitations and could be considered as a tool to detect the structural transition to a zigzag ordering. 
 By analyzing density evolution of longitudinal static structure factor and pair distribution function we have shown that driving force of zigzag ordering is caging phenomenon that in the case of hard-disk system is governed by excluded volume interaction with first and second neighbors and is of entropic origin.

\end{abstract}


\maketitle

In recent years there has been renewed interest~\cite{HicksPRL2018,HuMP2018,FuSM2017,NygardPRX2016,GodfreyPRE2016,YamchiPRE2015,Varga2011}
 on the properties of hard-core fluid 
confined to a channel of the width that does not exceed two particle diameters, so-called quasi-one-dimensional or q1D hard-core systems. The reasons are both 
basic and applied. The fundamental interest stems from existence of the exact analytical solutions for 1D~\cite{Tonks} and q1D~\cite{Barker1962,Kofke1993} systems and concerns the transformation of properties of these systems upon their progression into 2D and 3D directions. 
The practical interest 
concerns the possibility to use a simple q1D model to capture some particular features of more complex behavior of the confined fluids, e.g., to explain the single-file fluid properties in zeolite and carbon channels~\cite{Fois,Kofinger,Waghe}, microfluidic devices \cite{Mark} \textit{etc.} by treating the finite length axis 
as the pore width~\cite{Varga2011,Kofke1993}.

Computer simulations methods both Monte Carlo and molecular dynamics being a perfect tool to explore the structure, thermodynamics 
and dynamics of the systems with wide spectrum of continuous interparticle interactions are eligible to study the 
properties of hard-core fluids as well~\cite{Allen}. 
However only recently \cite{Hue15,Hue17} molecular dynamics (MD) simulations were exploited to 
find out how collective dynamics of bulk (3D) and confined (2D) hard-core systems behaves on 
different spatial scales~\cite{Rice}. More specifically, for the case of 2D HD fluid 
rather unexpectedly  the short-wavelength shear waves (SWSW) were found~\cite{Hue15} from well-defined 
peaks of the transverse current spectral functions. 
Interestingly, that SWSW while being absent at low densities were observed in the range of high densities by
showing particular features just before the freezing transition in 2D HD fluid~ \cite{Hue15}. 

The present paper deals with q1D HD system and there were few reasons that guided us into this study. 
First of all, it is strong correlation between emergence of SWSW and  the caging phenomenon~\cite{Trucket,adtpre2006} 
that has been noticed  in the case of 2D HD system~\cite{Hue15}.
Therefore, since caging inevitable will emerge in q1D HD system, it is quite naturally to expect the existence of SWSW in this case as well.  
If so, then it is not clear how collective modes, that were observed for 2D HD fluid, will be affected by an extra confinement, 
 i.e., how disk's reflection from channel walls will affect the longitudinal (L) and transverse (T) excitations.  
The 
latter are of particular interest since 
T-excitations 
are not present in 1D prototype of the q1D HD system.
Finally, among various consequences of caging the most intriguing is its role \cite{adtpre2006} in the freezing transition in 2D HD system,
that allows us to speculate on similar issue but now in the case of q1D HD system.

\begin{figure}[h]
\includegraphics[width=0.40\textwidth]{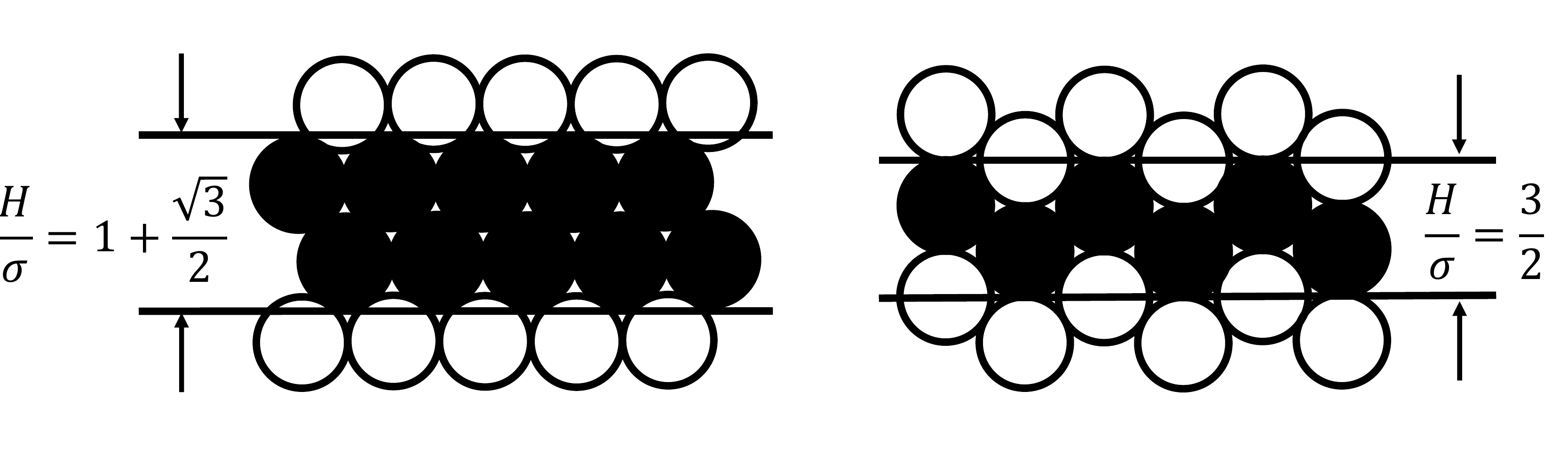}%
	
\includegraphics[width=0.48\textwidth]{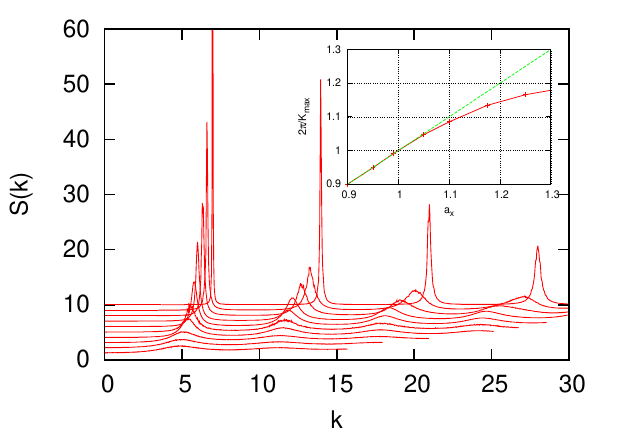}%
\caption{\textit{Top:} 
         Horizontally (left) and vertically (right) oriented triangular lattices and 
         zigzag ordering in q1D HD system.  
		\textit{Bottom:} Structure factor $\,S(k_x)\,$ of q1D HD system of the width $\,H/\sigma = 3/2\,$ for a  set of ten channel lengths $\,L\,$ considered in the present study; data for the shortest channel 
        $\,L/\sigma=180\,$ are at the top.
		Inset shows dependence of the inverse $\,2\pi/k_{\rm max}\,$ of  position of the first peak of $\,S(k_x)\,$ on inverse density parameter  $\,a_x=\rho^{-1}\,$.
	} \label{sk}
\end{figure}

The q1D HD system is modeled by placing  
$\,N\,$ hard-disks of diameter $\,\sigma\,$ into a rectangular box (channel) formed by two horizontal hard lines (walls) of length $\,L_x\equiv L\,$ that are separated in vertical $\,y\,$-direction by a distance (channel width) $\,L_y\equiv H<2\sigma\,$  such that disks cannot pass each other.
Taking into account that width $\,H=\sigma\,$ corresponds to 1D case, the range of channel widths $\,\sigma< H<2\sigma\,$
could be considered as a bridge between 1D and higher dimensions.
Among continuous variety of q1D HD systems in this range 
there are two when disk ordering at close packing is commensurate with 2D triangular lattice.
The cartoons of such q1D systems that correspond to $\,H/\sigma = 1 + \sqrt{3}/2\,$ and $\,H/\sigma = 3/2\,$ are shown 
on Fig.~\ref{sk}, where two triangular orderings 
correspond to 
horizontal and vertical orientations of triangular lattice that differ by an angle of 30 degrees.
In the present study we are considering q1D system of width fixed at $\,H/\sigma = 3/2\,$. 
The ends 
are open  in $\,x$-direction and periodic boundary conditions are employed.

The disk-disk and disk-wall interaction potentials are given by
\begin{equation}
  u(r_{\rm})=\left\{
\begin{array}{ll}
\infty, &  r<\sigma  \\
0, &  r\ge \sigma\,,
\end{array}
\right. \quad
u_{\rm w}(y_{\rm})=\left\{
\begin{array}{ll}
0, &  \sigma/2<y<\sigma  \\
\infty, &   \mbox{otherwise}\,,
\end{array}
\right. 
\end{equation}
%
%
%
where $\,r_{\rm}=|{\bf r}_{\rm i}-{\bf r}_j|\,$ is the distance between disks $i$ and $j$, 
while $\,y\,$ stands for the ordinate of each disk.

We employed the event driven molecular dynamics (MD) simulation algorithm \cite{Allen} in the canonical $NVT$ ensemble.
The main body of computer simulations were performed by using fixed collection of $\,N=200\,$ HD particles
To study the dependence of system properties on 
particle density, the simulation runs have been performed at different channel length $\,L\,$. 
Aiming to preserve condition 
$\,L>>H\,$,
the shortest  channel length was fixed at $\,L/\sigma=180\,$ while whole set of the studied q1D systems includes  another nine channel lengths, i.e.,
$\,L/\sigma=190,$ 198, 210, 220, 230, 250, 300, 350 and 400. 
To examine the system size effects 
simulation runs with $\,N=400\,$ particles still were performed.

Figure \ref{sk} shows simulation data for longitudinal  static structure factor $\,S(k_x)\,$ for all systems under study.
The former was  evaluated in a standard way \cite{HansenMcD} as instantaneous-time density-density correlator $\,S(k_x)=\langle n(-k_x)n(k_x)\rangle \,$
%
%
and concerns the changes in disks structuring upon changing the disk density.
One can see that for shortest channel, $\,L/\sigma=180\,$, the structure factor 
exhibits main  and few neighboring peaks like a sheared-out delta function that might be treated as being typical for distorted crystal state. 
As channels become longer, the shape of $\,S(k_x)\,$ becomes a typical of HD fluid at liquid and  gaseous state densities. 

It worth to remind that 
channel width $\,H/\sigma=3/2\,$ is commensurable with 2D triangular lattice. Consequently the zigzag arrangement, that eventually will appear in HD system under a narrow channel confinement~\cite{Varga2011}  and presumably could be associated with a solid state, in the present case will be just a fragment 
of the hexagonal ordering of HD particles  in a vertically oriented 2D triangular lattice. 
Therefore, rather convenient parameter to characterize the system under study  is the dimensionless linear number density $\,\rho=N\sigma/L\,$ that in the present case will be changing in the range from $\,\rho = 0.5\,$ till 1.1111;
the highest, i.e. close packing (cp) linear density in  crystalline zigzag arrangement is $\,\rho_{\rm cp}=1.1547$.
As for zigzag ordering of HD particles itself, it is evident that latter could appear at linear densities $\,\rho>1\,$ only, when there will  not be enough place for disks to align, i.e., when the distance between neighboring disks in $x$-direction will be less than disk diameter $\,\sigma$, or when the inverse density parameter $\,a_{x}=\rho^{-1}<1\,$. 

Following above discussion, we turn attention to a notable drift of the position $\,k_{\rm max}\,$ of the first maximum of $\,S(k_x)\,$ towards
smaller values of $\,k\,$ as disk density decreases.
In disordered systems, where the structural transition takes place, one can 
find out different slopes for the density dependence of position $\,k_{\rm max}\,$  on both sides of the transition~ \cite{Bry13b}. 
In the present case we plot (the inset of Fig.~\ref{sk})   the inverse of position of the first peak $\,2\pi/k_{\rm max}\,$ 
upon increase of the inverse density parameter $\,a_x\,$ that stands for an increase of the system length per particle.
Indeed, in the range of densities $\,\rho>1\,$,
the quantity $\,2\pi/k_{\rm max}\,$, i.e., average interparticle distance increases linearly  upon increasing the
system length. Such an observation indicates that these values of $k_{\rm max}$ could be the vectors of the same reciprocal lattice of the considered q1D HD system. On other hand, when density $\,\rho<1\,$, 
the increase of 
$\,2\pi/k_{\rm max}\,$ deviates from linear law that is expected for disordered  fluid systems.
In what follows, we will refer to the range of densities $\,\rho<1\,$ as  the fluid-like densities, while densities in the range  $\,\rho>1\,$ will be referred to as the solid-like densities.

\begin{figure}[h]
\includegraphics[width=0.48\textwidth]{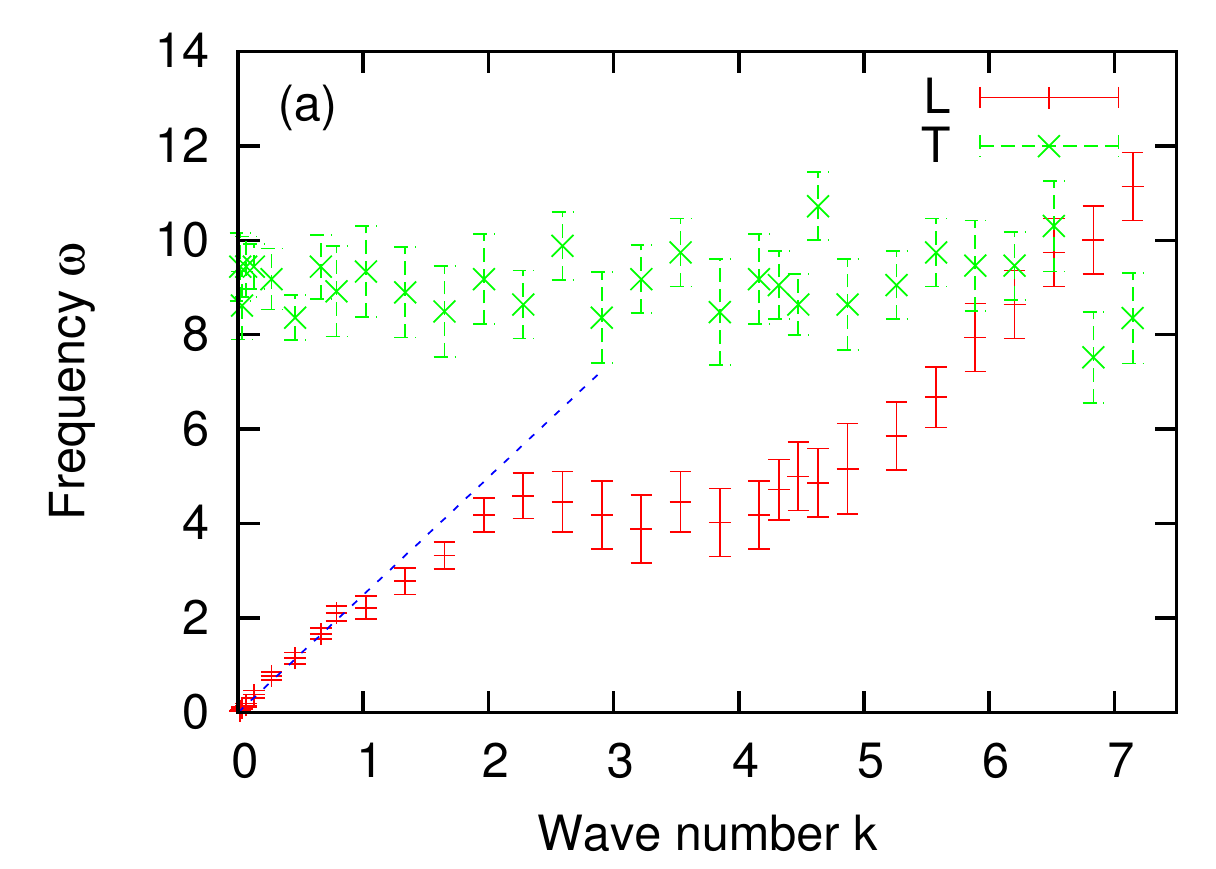}%

\includegraphics[width=0.48\textwidth]{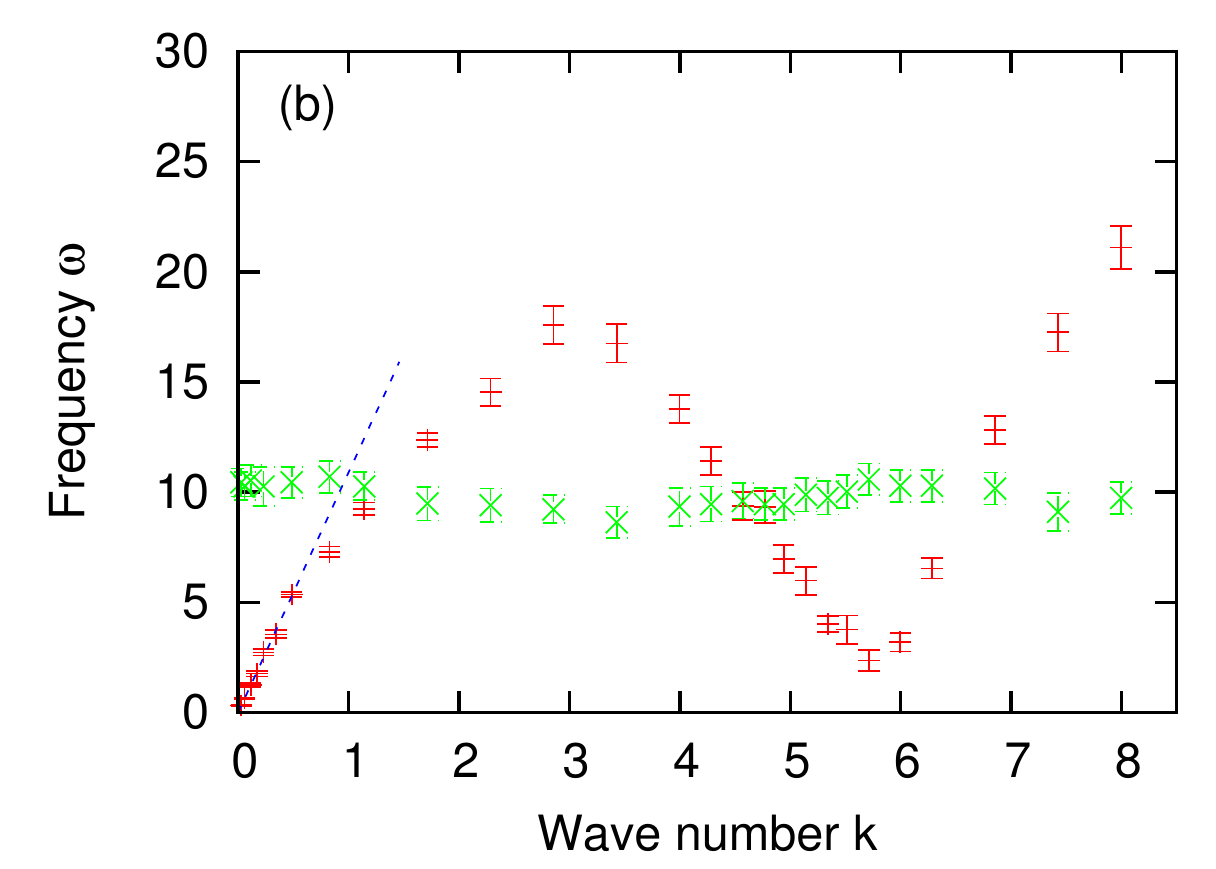}%
	\caption{Dispersion $\,\omega(k)\,$ of longitudinal (L) and transverse (T) excitations
		in q1D HD system of the width $\,H/\sigma=3/2\,$  in the range of fluid-like densities 
$\,\rho=0.5\,$ in part a) and  $\,\rho=0.9091\,$ in part b). 
	   	Dashed straight line at the small wavenumber values
		corresponds to hydrodynamic dispersion law. 
	} \label{disp1}
\end{figure}
%

\begin{figure}[h]
\includegraphics[width=0.48\textwidth]{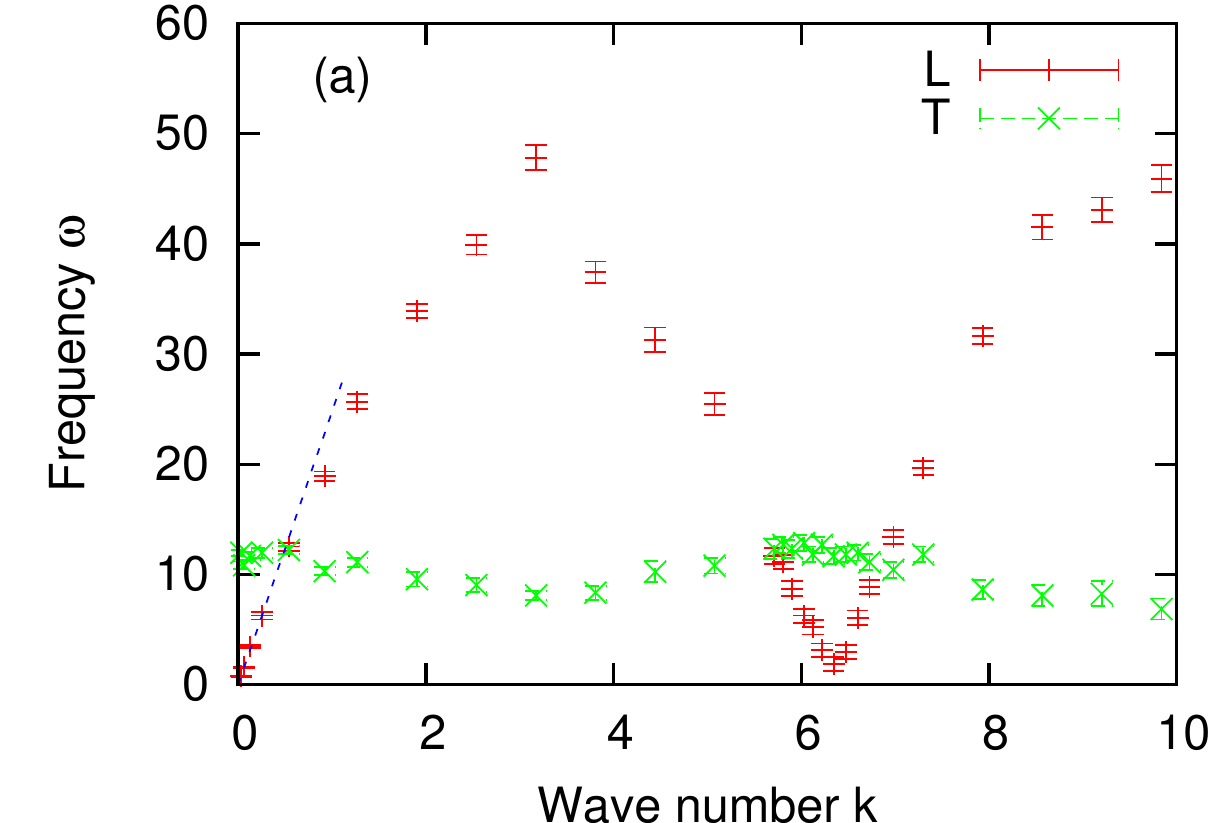}%

\includegraphics[width=0.48\textwidth]{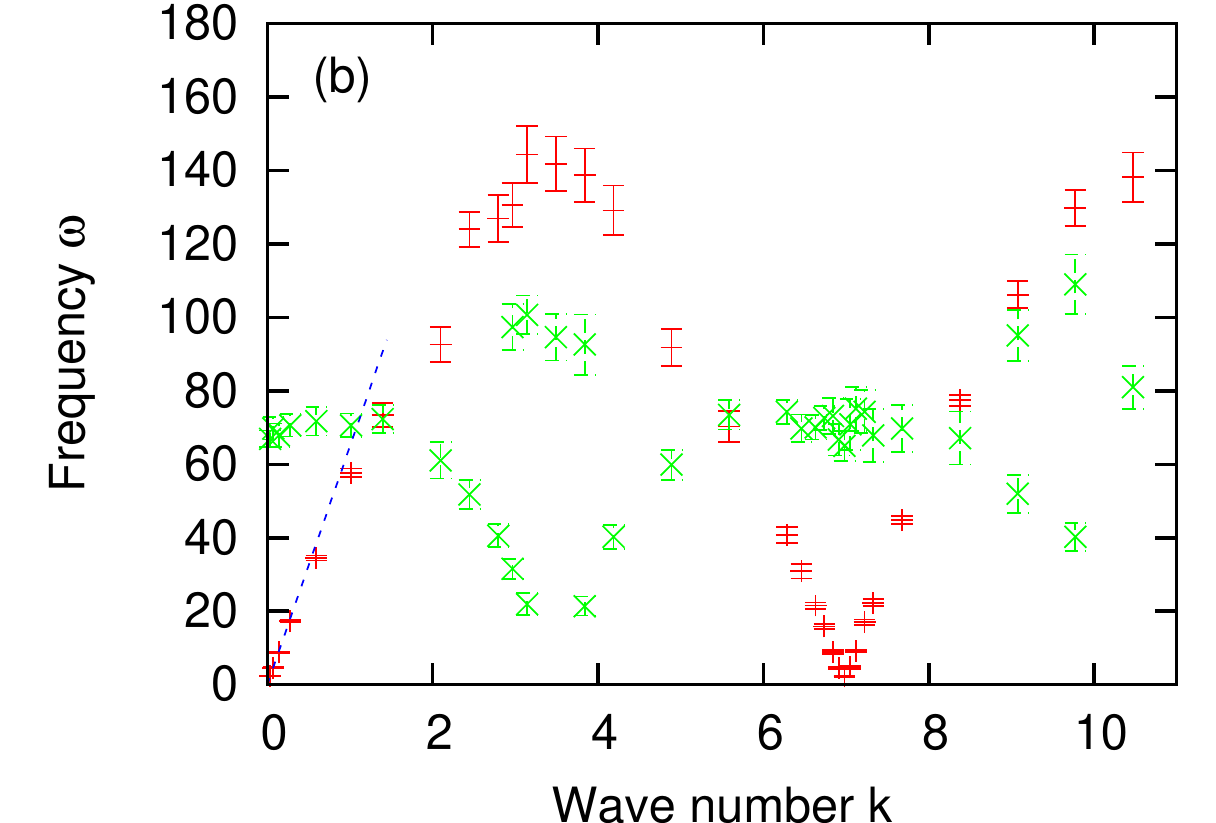}%
	\caption{The same as in Fig.~\ref{disp1}, but in the range of solid-like densities 
  $\,\rho=1.0101\,$ in part a) 	and   $\,\rho=1.1111\,$ in part b).
	} \label{disp2}
\end{figure}

Figures \ref{disp1} and \ref{disp2} show the dispersions of L and T collective excitations.
Both functions  $\,\omega^{\rm L}(k)\,$ and $\,\omega^{\rm T}(k)\,$ were obtained from the peak positions of respective current-current spectral functions
$\,C^{\rm L/T}(k,\omega)$ that in turn are time-Fourier transforms of the simulation data for 
current-current time correlation functions $\,F^{L/T}(k,t)$.  
Figure~\ref{disp1} presents data for the fluid-like density range.
At first glance  in  this case the dispersion of L-excitations 
in q1D HD system is
very similar to that already observed for a 2D HD system \cite{Hue15}. When density is low $\,\rho=0.5\,$, the dispersion $\omega^{\rm L}(k)$ only slightly deviates from being monotonic. But as soon as density increases
it shows well defined minimum  around wavenumber values $\,k\sim 6\,$ associated with position of the main peak of $\,S(k_x)\,$. 
The deviation from hydrodynamic dispersion law in the long-wavelength limit for both considered densities persists to be "negative". 
In contrast,  the dispersion of T-excitations 
is essentially different from similar in the case of 2D HD system \cite{Hue15}. Namely, for both densities we are observing rather flat
curve $\,\omega^{\rm T}(k)\,$ at frequency $\,\sim 10\,$ with a tendency towards higher values with increase of density. 
This kind of T-modes is induced by the reflections of disks from confining hard walls. 
In 2D HD system  the T-excitations  are of acoustic nature; they are absent 
at low densities and there was observed a long-wavelength propagation gap at higher densities.

Figure \ref{disp2} presents similar data but for the solid-like range of densities. 
As for L-excitations  we see the tendencies already observed in Fig.~\ref{disp1} under increase of density, i.e., the magnitudes of $\,\omega^{\rm L}(k)\,$ maxima are increasing while minima become deeper, reaching zero-frequency values at density $\,\rho=1.1111\,$ and being shifted towards larger wavenumber values $\,k\,$. 
Such behavior of $\,\omega^{\rm L}(k)\,$  resembles one for the ordered solids and, in general, could be interpreted as consequence of the 
ordering that emerges in a squeezed q1D HD system.

The dispersion of T-excitations in Fig.~\ref{disp2}  does not show notable changes too when the density was changing from fluid-like to solid-like at $\,\rho=1.0101\,$.
However, it does show dramatic changes  at density $\,\rho=1.1111\,$. Namely, (i) there is a sharp increase of frequency 
$\,\omega^{\rm T}\,$ up
to $\,\sim 70\,$; (ii) the dispersion  curve itself 
exhibits bubble-like shape by splitting  on low- and high-frequency branches
in the range 
of the maximum of dispersion $\,\omega^{\rm L}(k)\,$ that
implies possibility of L-T excitation coupling on atomic scale in a squeezed 
and ordered q1D HD system.

\begin{figure}[h]

  	\includegraphics[width=.48\textwidth]{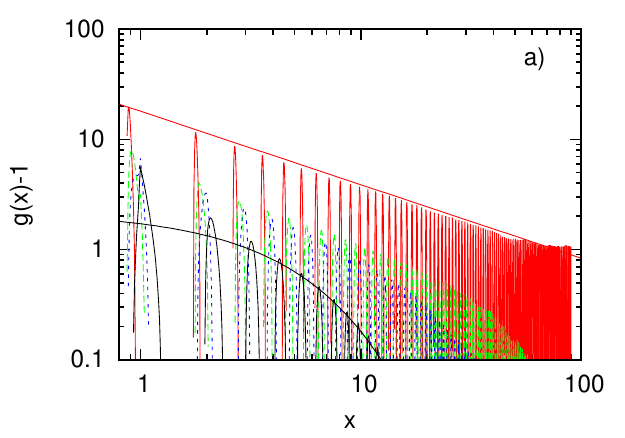}

	\includegraphics[width=.48\textwidth]{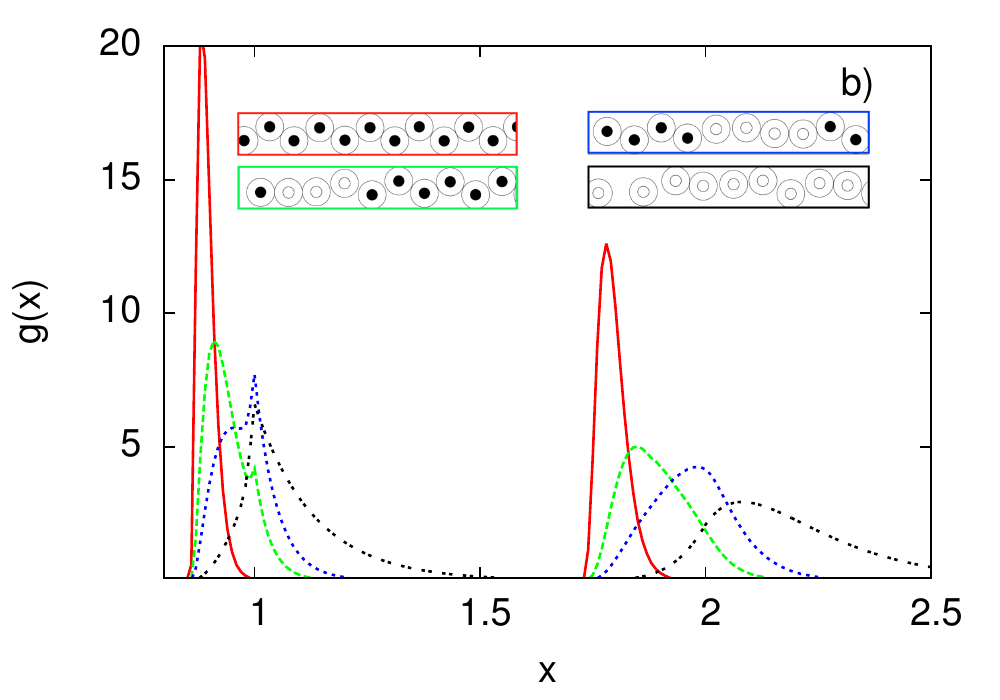}

	\caption{Pair distribution function $\,g(x)\,$ of q1D HD 
		system of the width $\,H/\sigma=3/2\,$ at four densities 
		$\rho=1.1111$, 1.0526, 1.0101 and 0.9091 from the top to the bottom on part a)  and from the left to the right on part b).
		\textit{Top:} Log-log view of the positional correlations given by the decay of the peaks of $\,g(x)-1\,$. Positional order for 
        density $\,\rho=1.1111\,$ 	is algebraic $\,\sim x^{-2/3}\,$, while  positional order for 
        density $\,\rho=0.9091\,$ is short-ranged with exponential decay $\,\sim\exp(-x/4)\,$.
		\textit{Bottom:} First and second peaks of $\,g(x)\,$ only.
		Inset shows representative snapshots of the disk configurations at each discussed density; snapshot boxes and corresponding $\,g(x)\,$ curves are of the same color
		while filled circles indicate caged disks.
	} \label{rdf}
\end{figure}

The deeper insight into onset of presumably solid-like zigzag ordering associated with densities $\,\rho > 1\,$ could be obtained from 
Fig.~\ref{rdf} that presents simulation data for pair distribution function $\,g(x)\,$ defined as conventional radial distribution function $\,g(r)\,$, but just along  $\,x$-direction only. By looking first on the snapshots of disk configurations
in Fig.~\ref{rdf}b it follows that at density $\,\rho=1.1111\,$ zigzag ordering indeed is taking place. 
Moreover, it is confirmed by positions of the first and second 
peaks of  $\,g(x)\,$ that are located at $\,x=0.880\pm 0.005\,$ and $\,1.780\pm 0.005$, respectively. 
The latter means that at density $\,\rho=1.1111\,$ the distance between centers of two nearest neighbors of each disk in the system is less than $\,2\sigma\,$, 
i.e., each disk is caged. 
We note, that for zigzag ordering at close packing
these peak positions should be at  $\,x=\sqrt{3}/2\approx 0.866\,$ and $\,\sqrt{3}\approx 1.732$, respectively. 

By looking carefully on the cartoon in Fig.~\ref{sk} it can be found that two nearest neighbors of each disk in q1D HD system in fact are 
the second neighbors with respect one to another in the case of 2D HD system. The shortening of distance between centers of these two disks causes shoulder 
on the second peak of $\,g(r)\,$ in 2D HD system, reflecting emergence of the caging and has been recognized \cite{Trucket} as structural precursor to freezing  in 2D and 3D hard-core systems.   In the case of q1D HD system corresponding shoulder is observing on the first peak of $\,g(x)\,$ 
for densities $\,\rho>1\,$ and by analogy could be referred  to as structural precursor to zigzag ordering.  

Obviously, caging is absent for densities $\,\rho<1\,$. By emerging in the range of densities $\,\rho>1\,$ it becomes the reason of linear dependence of 
$\,2\pi/k_{\rm max}\,$ upon increase of the inverse density parameter 
in Fig.~\ref{sk}. However, such a partial caging only slightly affects the decay of positional correlations for densities $\,1<\rho<1.1111\,$; by remaining short-ranged, the former already 
deviates from an exponential decay that takes place for  fluid-like densities (see two intermediate curves on Fig.~\ref{rdf}a and one for density $\,\rho=0.9091\,$ that decays as $\,\sim \exp(-x/4)\,$). 
Only at density $\,\rho=1.1111\,$ positional correlations show algebraic decay $\,\sim x^{-2/3}\,$ indicating the quasi-long-ranged positional order, i.e., at this density q1D HD system could be already in solid phase or very close to the solid phase. 

Summarizing, we performed MD simulation studies of the q1D system composed of hard disks of diameter $\,\sigma\,$ confined between two parallel lines a distance $\,H/\sigma=3/2\,$ apart. This system is of particular interest since at disk close packing
it allows for zigzag ordering commensurable with vertically oriented 2D triangular crystalline solid.  There are computer simulation studies that already have discussed the possibility of zigzag-like ordering in q1D HD systems, including one characterized by the same width parameter $\,H\,$ \cite{Varga2011}.
However, to the best of our knowledge there has not been reported microscopic physical property that is sensitive and/or could point out to such a structural transition. 

In the present study we presented simulation data for dispersion of T-excitations that could fulfill that gap.
Obviously, any T-excitations could exist in limiting case of 
classical 1D gas~\cite{Tonks}. From other hand, an existence of the short-wavelength shear waves (SWSW) in 3D and 2D counterparts of the q1D HD system was found recently~\cite{Hue15,Hue17} and
was attributed to the caging phenomenon  that among other is responsible for freezing transition in these systems.  
We have shown that caging does exist in 1D HD system as well, and is featured by the positions of first two peaks of $\,g(x)\,$ which 
unambiguously tell us that at density $\,\rho\ge 1.1111\,$ any two neighboring disks at the same channel wall do not permit for the disk in between them to pass through.
However, in contrast to 2D HD case, caging in q1D HD system is not the origin of T-excitations; those already exist at density as low as $\,\rho=0.5\,$ 
because of reflection of uncaged disks from confining walls due to excluded volume interaction. Instead of this, caging in q1D HD system does initiate the split of dispersion $\,\omega^{\rm T}(k)\,$ into two frequency branches associated with the movement of caged disks.
In our particular q1D HD system we found that zigzag ordering already exists at density $\,\rho=1.1111\,$ when distance between neighbors in the system is  $\,0.9\sigma\,$, while for zigzag crystalline solid it is $\,\sqrt{3}\sigma/2\sim 0.866\sigma\,$.
Finally,  shoulder on the first peak of $\,g(x)\,$ 
could be considered as structural precursor to zigzag ordering, by analogy with 2D and 3D hard core systems~\cite{Trucket}.   \\


AH thanks the support of Conacyt M\'exico during the sabbatical leave and valuable discussions 
with Dra. Karen Volke and Dr. Alejandro Vasquez of the IFUNAM.

%

%
\end{document}